\documentclass[twocolumn,showpacs,preprintnumbers,amsmath,amssymb,times,pre]{revtex4}

\usepackage{graphicx}% Include figure files
\usepackage{dcolumn}% Align table columns on decimal point
\usepackage{bm}% bold math
\usepackage{amsmath}

\begin{document}

\title{Role of aspiration-induced migration in cooperation}

\author{Han-Xin Yang$^{1}$}\email{hxyang@mail.ustc.edu.cn}

\author{Zhi-Xi Wu$^{2}$}\email{zhi-xi.wu@physics.umu.se}
\author{Bing-Hong Wang$^{1,3}$}\email{bhwang@ustc.edu.cn}

\affiliation{$^{1}$Department of Modern Physics, University of
Science and Technology of China, Hefei 230026,
China\\$^{2}$Department of Physics, Ume\aa\, University, 90187
Ume\aa, Sweden\\$^{3}$The Research Center for Complex System
Science, University of Shanghai for Science and Technology and
Shanghai Academy of System Science, Shanghai, 200093 China }

\begin{abstract}
Both cooperation and migration are ubiquitous in human society and
animal world. In this paper, we propose an aspiration-induced
migration in which individuals will migrate to new sites provided
that their payoffs are below some aspiration level. It is found that
moderate aspiration level can best favor cooperative behavior. In
particular, moderate aspiration level enables cooperator clusters to
maintain stably and expand efficiently, whereas induces defector
clusters to disintegrate, thus promoting the diffusion of
cooperation among population. Our results provide new insights into
understanding the role played by migration in the emergence of
cooperative behavior.

\end{abstract}

\date{\today}


\maketitle

Cooperation is fundamental to biological and social systems. Many
important mechanisms have been considered for studying the
cooperative behavior, such as costly
punishment~\cite{punishment1,punishment2}, reputation~
\cite{reputation1,reputation2} and social diversity
~\cite{diversity1,diversity2,diversity3}. As is well-known,
migration is a common and essential feature present in animal world
and human society. For example, every year millions of animals
migrate in the savannas of Africa, and every day thousands of people
travel among different countries. Recently, the role of migration
has received much attention in the study of evolutionary
games~\cite{migration1,migration2,migration3,success-driven,migration4,migration5,migration6}.

Migration can be in a random-walk way. For example, Vainstein $et$
$al.$ studied the case in which individuals are located on the sites
of a two-dimensional regular lattice and each individual makes an
attempt to jump to a nearest neighboring empty site chosen randomly
with some probability~\cite{migration3}, Meloni $et$ $al.$ consider
the case in which individuals are situated on a two-dimensional
plane and each individual moves to a randomly chosen position with
certain velocity~\cite{migration5}. Besides, the direction of
migration can be payoff biased, that is, individuals choose the
destination of migration according to payoff. For example, Helbing
$et$ $al.$ proposed a success-driven migration mechanism in which
individuals will move to the sites with highest estimated
payoffs~\cite{success-driven}, Boyd $et$ $al.$ divided individuals
into different subpopulations, the fraction of individuals in
subpopulation $i$ who leave and join subpopulation $j$ depends on
the payoff difference between two subpopulations~\cite{migration6}.

It is noted that in many real-life situations, individuals have to
migrate when they cannot find enough resources for living. For
example, animals will migrate to other places if they cannot find
enough food in the current habitats. Based on such consideration, we
propose an aspiration-induced migration model, wherein each
individual plays with all its neighbors and accumulates payoffs
correspondingly; An individual will move to another place if its
payoff is lower than the aspiration level, and stay otherwise in the
current location. Here the aspiration level can be understood as the
minimum living standard for each individual. Considering limited
information of the individuals, we assume that, migrants choose new
places in a random way.

We use the famous prisoner's dilemma game (PDG)~\cite{prisoner's
dilemma} to carry out our researches. In the PDG played by two
players, each of whom chooses one of two strategies, cooperation or
defection. They both receive payoff $R$ upon mutual cooperation and
$P$ upon mutual defection. If one defects while the other
cooperates, cooperator receives $S$ while defector gets $T$. The
ranking of the four payoff values is: $T>R> P>S$. Following common
practice~\cite{spatial}, we set $T=b$ ($>1$), $R=1$, and $P=S=0$,
where $b$ represents the temptation to defect.

We assume prisoner's dilemma players on a square with periodic
boundary conditions and $L \times L$ sites, which are either empty
or occupied by one individual. Initially, an equal percentage of
strategies (cooperators or defectors) is randomly distributed among
the population. Individuals are updated asynchronously, in a random
sequential order. The randomly selected individual plays against
individuals sitting on four neighboring nodes (the von Neumann
neighborhood), collecting the payoff from the combats. The
individual compares its total payoff with its direct neighbors and
changes strategy following the one (including itself) with the
highest payoff. Before updating strategy, an individual decides
whether to stay at or leave its current site. Individual stays in
current site if its payoff reaches or exceeds its aspiration level,
otherwise it moves to a randomly chosen empty site within its four
neighboring sites. To avoid isolated case, we assume that an
isolated individual makes mandatory move.

Following previous study~\cite{aspiration}, the aspiration level
$P_{ia}$ for an individual $i$ is defined as $P_{ia}=k_{i}A$, where
$k_{i}$ is the number of neighbors of $i$ and $A$ is a control
parameter ($A$ is the same for all individuals). This definition is
based on the following consideration: In real life costless
relationships are rare; To maintain social relationship, some cost
are required; We assume for simplicity that the needed cost is the
same for each link. Thus it is reasonable to assume the aspiration
level to be proportional to the number of neighbors.

\begin{figure}
\begin{center}
 \scalebox{0.82}[0.82]{\includegraphics{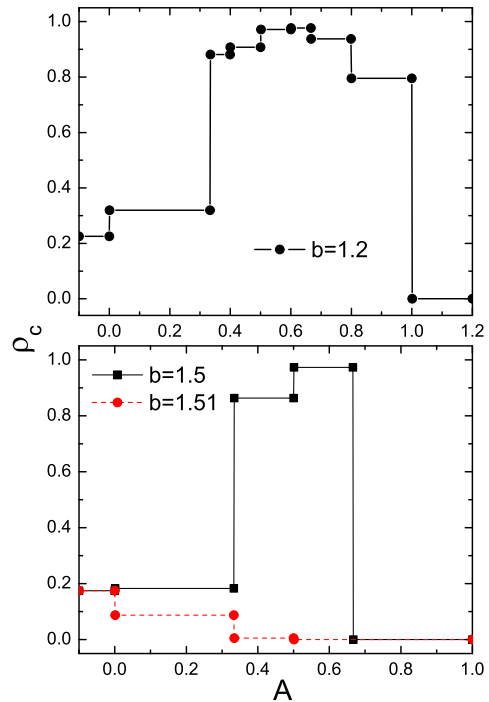}}
\caption{(Color online) Fraction of cooperators $\rho_{c}$ as a
function of the aspiration level $A$ for different values of the
temptation to defect $b$. The simulations are for $100 \times 100$
grids with the fraction of occupied sites $f=0.5$. The equilibrium
fraction of cooperators results from averaging over 2000 time steps
after a transient period of 20000 time steps. Each time step
consists of on average one strategy-updating event of all the
individuals. Results are averaged over 100 different realizations.}
\label{aspiration}
\end{center}
\end{figure}

Figure~\ref{aspiration} shows the fraction of cooperators $\rho_{c}$
as a function of the aspiration level $A$ for different values of
the temptation to defect $b$ when the fraction of occupied sites
$f=0.5$. One can see that, $\rho_{c}$ exhibits discontinuous phase
transition with varying $A$ and $\rho_{c}$ is the same between two
nearby phase transition points. The value of phase transition point
can be determined by the average payoff of an individual (total
payoff divided by the number of neighbors), which may be 0, 1/2,
$b/2$, 1/3, 2/3, $b/3$, $2b/3$, 1, $b$ (here we exclude the isolated
case in which individual makes mandatory move and four-neighbors
case in which individual cannot move). Taking $b=1.5$ as example,
the phase transition values of $A$ are 0, 1/3, 1/2, 2/3 respectively
(here 0.75, 1 and 1.5 are excluded since $\rho_{c}=0$ for $A>2/3$
when $b=1.5$)  \cite{note}.

From Fig.~\ref{aspiration}, one can also find, for a fixed value of
the temptation to defect $b$, there exists an optimal region of $A$,
leading to the highest cooperation level. For $b=1.2$ and $b=1.5$,
the optimal region of $A$ is (0.6, 2/3] and (0.5,2/3] respectively,
indicating that moderate aspiration level best favors cooperation.
For $b=1.51$, the optimal region of $A$ is ($-\infty$, 0], in which
all individuals do not move. This is because, according to the
analysis in Ref. \cite{b}, compared with never-move case, migration
makes it easier for defectors to invade cooperator clusters when
$b>1.5$.

\begin{figure}
\begin{center}
 \scalebox{0.75}[0.75]{\includegraphics{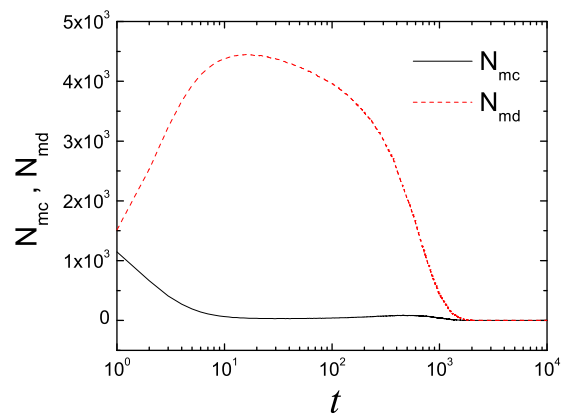}} \caption{(Color online) The number of mobile cooperators
 $N_{mc}$ and mobile defectors $N_{md}$ as a function of time step $t$ on $100 \times 100$ grids with
50\% empty sites. $b=1.5$, $A=0.6$. Results are averaged over 100
different realizations.} \label{time}
\end{center}
\end{figure}

How to understand moderate aspiration level best promotes
cooperation when $b\leq1.5$? It has been known that in spatial
games, cooperators can survive by forming clusters, in which the
benefits of mutual cooperation can outweigh losses against
defectors, thus enable cooperation to be
maintained~\cite{spatial,cluster}. For low aspiration level, most
individuals do not move. Consequently, cooperator- and
defector-clusters coexist and keep almost unchanged in the
stationary state, inhibiting the dispersal of cooperation among
population. On the contrary, for high aspiration level, most
individuals move. Due to the frequent change of neighbors,
cooperators can not form clusters to resist the invasion of
defectors. As a result, cooperators are doomed to extinct, analogous
to the situation arising in the well-mixed population.

\begin{figure*}
\begin{center}
 \scalebox{0.85}[0.85]{\includegraphics{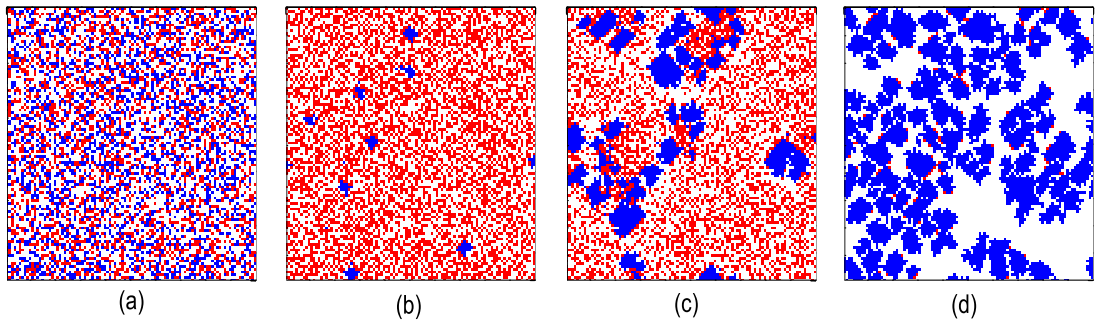}} \caption{(Color online) Snapshots of typical distributions of
 cooperators (blue) and defectors (red) at different time steps $t$ for $b=1.5$ and $A=0.6$. The simulations are for $100 \times 100$ grids with
50\% empty sites (white). (a) $t=1$, $\rho_{c}(1)=0.5$, (b) $t=16$,
$\rho_{c}(16)=0.0298$, (c) $t=370$, $\rho_{c}(370)=0.302$,
 and (d) $t=3000$, $\rho_{c}(3000)=0.976$.}
 \label{map}
\end{center}
\end{figure*}

For moderate aspiration level, cooperators can form stable clusters
since high benefits of mutual cooperation insure them to stay in
cooperator clusters, whereas defectors avoid gathering together
because the payoffs of mutual defection are low. Figure \ref{time}
shows that, during the process of evolution, the number of mobile
defectors $N_{md}$ is much larger than mobile cooperators $N_{mc}$
when $b=1.5$, $A=0.6$, indicating moderate aspiration level enables
cooperator clusters to be sustained whereas induces defector
clusters to be disintegrated. A mobile defector would change to
cooperator if it touches cooperator cluster and encounters a
cooperator who has the highest payoff among defector and its
neighbors (this situation is likely to occur since cooperator
clusters usually obtain high payoffs). Thus, for moderate aspiration
level, cooperator clusters not only be able to maintain, but also
expand due to the existence of migration.

\begin{figure}
\begin{center}
 \scalebox{0.75}[0.75]{\includegraphics{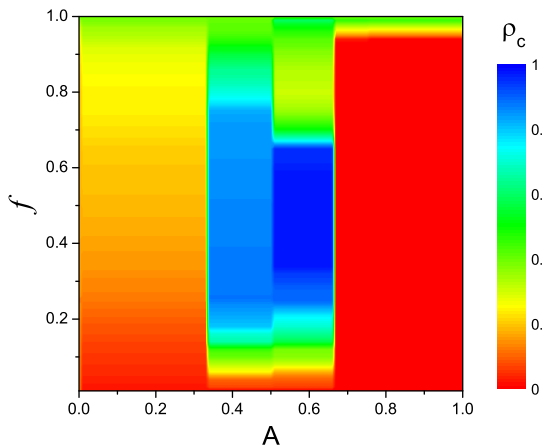}} \caption{(Color online) The color code shows the fraction of
cooperators $\rho_{c}$ as a function of the aspiration level $A$ and
the fraction of occupied sites $f$ on $100 \times 100$ grids. The
temptation to defect $b=1.5$. Results are averaged over 100
different realizations.} \label{color code}
\end{center}
\end{figure}

\begin{figure*}
\begin{center}
 \scalebox{0.82}[0.82]{\includegraphics{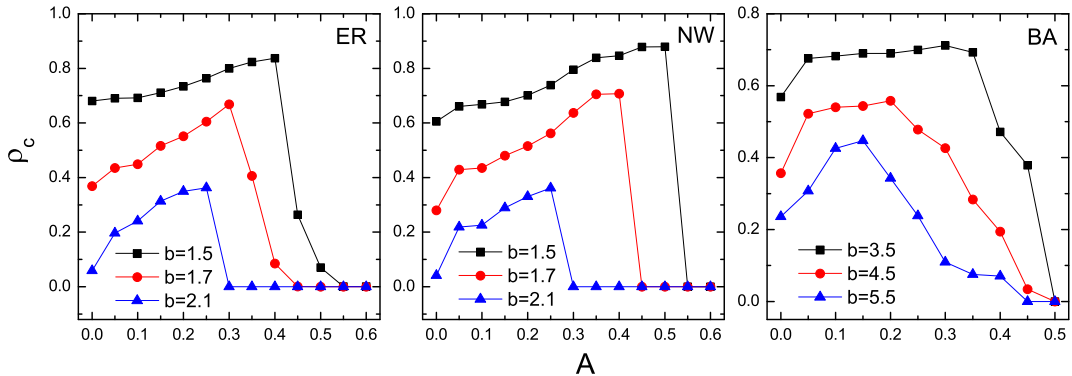}} \caption{(Color online) Fraction of cooperators $\rho_{c}$ as
a function of the aspiration level $A$ for different values of $b$
when individuals are situated on: ER random graphs (left panel), NW
small-world networks (middle panel) and BA scale-free networks
(right panel). All networks are of $10^{4}$ nodes with 50\% empty
nodes. Average connectivity of ER, NW and BA networks is 10, 8, 6
respectively. The equilibrium fraction of cooperators results from
averaging over 2000 time steps after a transient period of 20000
time steps. Results are averaged over 100 different realizations.
 } \label{network}
\end{center}
\end{figure*}

To intuitively understand how moderate aspiration level affects the
evolution of cooperation, we plot the distribution of cooperators
and defectors on a square lattice at different time steps $t$ for
$b=1.5$ and $A=0.6$. Initially ($t=1$), cooperators and defectors
are randomly distributed with the same probability on the square
lattice [see Fig. \ref{map}(a)]. From Fig. \ref{map}(b), we can see
that cooperators and defectors are quickly clustered respectively
($t=16$), and the density of cooperators at this moment is lower
than the initial state because cooperators are exposed to much
attack of defectors before the formation of steady cooperator
clusters. As time step $t$ increases, cooperator clusters expand and
defector clusters shrink [see Fig. \ref{map}(c)]. Finally,
cooperators take over the population and defectors only dispersedly
survive nearby cooperator clusters [see Fig. \ref{map}(d)],
demonstrating that moderate aspiration level can effectively impulse
the collapse of defector clusters.

The fraction of occupied sites $f$ also affects the evolution of
cooperation. Figure \ref{color code} shows the fraction of
cooperators $\rho_{c}$ as a function of the aspiration level $A$ and
the fraction of occupied sites $f$ together when the temptation to
defect $b=1.5$. From Fig. \ref{color code}, one can see that, the
optimal region of $A$ corresponding to the highest cooperation level
changes as $f$ varies. For example, the optimal region of $A$ is
(0.5,2/3] and (1/3,0.5] for $f=0.5$ and $f=0.8$ respectively.
Besides, one can find that, for a fixed value of $A$, $\rho_{c}$
varies as $f$ changes. For $A\leq1/3$ and $A>2/3$, $\rho_{c}$
increases as $f$ increases and $f=1$ corresponds to the maximum
$\rho_{c}$. For $1/3<A\leq2/3$, there exists an intermediate value
of $f$, leading to the highest cooperation level.

There is much current interest in studying evolutionary games on
various networks \cite{spatial 2,spatial 3,spatial 4,spatial
5,spatial 6,spatial 7,spatial 8}. It has been found that spatial
structure greatly affects the evolution of cooperation \cite{spatial
2}. As a natural extension, we now consider the aspiration-induced
migration on Erd\"{o}s-R\'{e}nyi random graphs (ER) \cite{ER},
Newman-Watts small-world networks (NW) \cite{NW} and
Barab\'{a}si-Albert scale-free networks (BA) \cite{BA}. In Fig.
\ref{network}, we report the fraction of cooperators $\rho_{c}$ as a
function of the aspiration level $A$ for different values of the
temptation to defect $b$. It is interesting to find that, for a
fixed value of $b$, there exists an optimal $A$ corresponding to the
maximum $\rho_{c}$ for all three networks.

In summary, we have incorporated an aspiration-induced migration
mechanism to the evolutionary prisoner's dilemma game. An individual
would migrate if its payoff is lower than the aspiration level. We
find that, for individuals locating on square lattice and the
temptation to defect $b<1.5$, there exists an optimal range of the
aspiration level, leading to the maximum cooperation level. We
explain such phenomenon by investigating the evolution of
cooperator- and defector-clusters. Moderate aspiration level induces
cooperator clusters to expand and defector clusters to disintegrate,
thus promoting the diffusion of cooperation among population. We
also study the effect of the fraction of occupied sites $f$ on
cooperation. Furthermore, studies of aspiration-induced migration
model on random graphs, small-world networks and scale-free networks
show that, there exists an optimal value of the aspiration level
that can best favor cooperative behavior for individuals situated on
these networks. Finally we have checked that our conclusions are
robust with respect to using different strategy updating rules, such
as Fermi updating rule \cite{Fermi1, Fermi2} and the finite
population analogue of the replicator dynamics \cite{spatial 2}.

We thank Petter Holme and Wen-Xu Wang for useful discussions. This
work is funded by the National Basic Research Program of China (973
Program No. 2006CB705500), the National Natural Science Foundation
of China (Grant Nos. 10975126, 10635040), the Specialized Research
Fund for the Doctoral Program of Higher Education of China (Grant
No. 20093402110032). Z.X.W. acknowledges the support from the
Swedish Research Council.


\begin{references}
\bibitem{note}
Note that the phase transition points disappear when $f\simeq1$.
\end{references}
\end{document}